\documentclass[superscriptaddress,aps,12pt,letterpaper,preprintnumbers,amsmath,amssymb,prd,nofootinbib]{revtex4-1}

\usepackage{graphicx}
\usepackage{epstopdf}
\usepackage{dcolumn}
\usepackage{bm}
\usepackage{color}
\usepackage{amsmath}
\usepackage{mathtools}
\usepackage{xfrac}

\usepackage{feynmp}
\usepackage[colorlinks]{hyperref}
\hypersetup{
     colorlinks   = true,
     citecolor    = red,
	linkcolor=blue
}

\usepackage[T1]{fontenc}
\usepackage{textcomp}
\usepackage{setspace}

\begin{document}

\newcommand{\KEV}{ {\rm keV} }
\newcommand{\MEV}{ {\rm MeV} }
\newcommand{\GEV}{ {\rm GeV} }
\newcommand{\TEV}{ {\rm TeV} }
\newcommand{\1}{\mbox{1}\hspace{-0.25em}\mbox{l}}
\newcommand{\headline}[1]{\noindent{\bf #1}}

\def\diag{\mathop{\rm diag}\nolimits}
\def\Spin{\mathop{\rm Spin}}
\def\SO{\mathop{\rm SO}}
\def\O{\mathop{\rm O}}
\def\SU{\mathop{\rm SU}}
\def\U{\mathop{\rm U}}
\def\Sp{\mathop{\rm Sp}}
\def\SL{\mathop{\rm SL}}
\def\tr{\mathop{\rm tr}}
\def\mpl{M_{\rm Pl}}

\def\IJMP{Int.~J.~Mod.~Phys. }
\def\MPL{Mod.~Phys.~Lett. }
\def\NP{Nucl.~Phys. }
\def\PL{Phys.~Lett. }
\def\PR{Phys.~Rev. }
\def\PRL{Phys.~Rev.~Lett. }
\def\PTP{Prog.~Theor.~Phys. }
\def\ZP{Z.~Phys. }

\def\dd{\mathrm{d}}
\def\ff{\mathrm{f}}
\def\BH{{\rm BH}}
\def\inf{{\rm inf}}
\def\ev{{\rm evap}}
\def\eq{{\rm eq}}
\def\Mpl{M_{\rm pl}}
\def\GeV{{\rm GeV}}
\def\TeV{{\rm TeV}}

\newcommand{\SM}{{\rm SM}}
\newcommand{\DM}{{\rm DM}}
\newcommand{\RH}{{\rm RH}}
\newcommand{\QCD}{{\rm QCD}}
\newcommand{\GUT}{{\rm GUT}}

\newcommand{\Red}[1]{\textcolor{red}{#1}}
\newcommand{\HL}[1]{\textcolor{blue}{\bf HL: #1}}
\newcommand{\TYL}[1]{\textcolor{cyan}{\bf TYL: #1}}
\newcommand{\KH}[1]{\textcolor{red}{\bf KH: #1}}
\newcommand{\add}[1]{\textcolor{green}{#1}}

\graphicspath{{Figures/}}

\title{
GUTzilla Dark Matter
}

\author{Keisuke Harigaya}
\author{Tongyan Lin}
\author{Hou Keong Lou}
 \affiliation{Department of Physics, University of California, Berkeley, California 94720, USA}
\affiliation{Theoretical Physics Group, Lawrence Berkeley National Laboratory, Berkeley, California 94720, USA}
\affiliation{Kavli Institute for the Physics and Mathematics of the Universe (WPI),The University of Tokyo Institutes for Advanced Study, The University of Tokyo, Kashiwa, Chiba 277-8583, Japan}

\begin{abstract}
Motivated by gauge coupling unification and dark matter, we present an extension to the Standard Model where both are achieved by adding an extra new matter multiplet. Such considerations  lead to a Grand Unified Theory with very heavy WIMPzilla dark matter, which has mass greater than $\sim 10^7$ GeV and must be produced before reheating ends. Naturally, we refer to this scenario as GUTzilla dark matter. Here we present a minimal GUTzilla model, adding a vector-like quark multiplet to the Standard Model. Proton decay constraints require the new multiplet to be both color and electroweak charged, which prompts us to include a new confining $SU(3)$ gauge group
that binds the multiplet into a neutral composite dark matter candidate.
Current direct detection constraints are evaded due to the large dark matter mass; meanwhile, next-generation direct detection and proton decay experiments will probe much of the parameter space. 
The relic abundance is strongly dependent on the dynamics of the hidden confining sector, and we show that dark matter production during the epoch of reheating can give the right abundance.
\end{abstract}

\date{\today}

\maketitle


\section{Introduction}
\label{sec:intro}

Grand unified theories (GUTs)~\cite{Georgi:1974sy, Georgi:1974yf} are one of the most attractive and well-studied scenarios for physics beyond the Standard Model (SM). With just the particle content of the SM, the three gauge couplings run tantalizingly close to one another at around $10^{15}$~GeV. However, they do not meet at a single scale. The possibility of a GUT thus motivates additional new matter below the GUT scale, which can modify the running and allow unification. In principle, such new matter may be present anywhere above the weak scale up to the GUT scale, and there are limitless possibilities.

On the other hand, the existence of dark matter (DM) provides one of the strongest signs of physics beyond the SM. The existing searches for DM have dominantly focused on weak-scale thermal relics within the  Weakly Interacting Massive Particle (WIMP) paradigm \cite{Lee:1977ua}; however, the lack of definitive signals from (in)direct detection experiments \cite{Akerib:2015rjg,Ackermann:2015zua} and at the Large Hadron Collider (LHC) \cite{Khachatryan:2014rra,Aad:2015zva} have placed increasingly stringent constraints on WIMP models.  Therefore, it is prudent to re-examine our theory assumptions and explore alternative DM beyond the WIMP, including DM at much higher mass scales.

In this paper, we propose an extension to the SM that gives both gauge coupling unification and a very heavy DM candidate, with mass well above the weak scale. We extend the SM with an additional matter multiplet $\chi$, which is part of a larger split GUT multiplet. In order for $\chi$ to give successful gauge coupling unification, the multiplet must be both electroweak and color charged. Due to the color charge, we are led to consider
$\chi$ charged under an additional confining hidden gauge group.
The DM is then a composite state with electroweak interactions, which can evade direct detection bounds for masses above $10^7$ GeV. 

We refer to such GUT-motivated heavy DM as GUTzilla DM, by analogy with the \mbox{WIMPzilla} DM scenario~\cite{Kolb:1998ki, Chung:1998rq, Chung:2001cb, Feldstein:2013uha, Harigaya:2014waa}. In general, very heavy DM cannot be produced  thermally during a radiation-dominated era; if the DM were in thermal equilibrium, then a large annihilation rate would be required to avoid overclosing the universe, which runs into unitarity bounds for DM masses above the 100 TeV scale~\cite{Griest:1989wd}. 
Instead, in the WIMPzilla scenario, the relic abundance is set before the end of reheating. Then the relic abundance is naturally suppressed if the reheating temperature is smaller than the DM mass, and it is possible for the DM mass to span many orders of magnitude. Specifically, we consider DM production in inflaton direct decay and production from the SM thermal bath during the reheating epoch.

Other models that accomodate both unification and DM with extensions to the SM have been studied before~\cite{Mahbubani:2005pt,Ibe:2009gt,Aizawa:2014iea, Antipin:2015xia}, including $SO(10)$ unification~\cite{Arbelaez:2015ila, Nagata:2015wna}. These models typically require multiple new particles and hierarchies of scales. Here, we will consider the 
scenario with only one new hierarchy associated with the $\chi$ multiplet. The $\chi$ multiplet will invariably be part of a split GUT multiplet, and we will simply assume that the splitting is accomplished by a minimal amount of fine-tuning.
We will comment on a possible connection between the fine-tuning and the anthropic principle later.

Our paper is structured as follows: 
In Sec.~\ref{sec:coupling}, we outline the requirements on the new matter multiplet $\chi$  such that gauge coupling unification is achieved; we show that in order to satisfy proton decay constraints, $\chi$ needs to have both electroweak and color charge. A viable DM candidate would then need to be a composite state composed of the $\chi$, which we assume is the result of a new confining gauge interaction. In Sec.~\ref{sec:model}, we construct a minimal model of GUTzilla DM, where $\chi$ is a fundamental of a confining $SU(3)_H$. For simplicity, our discussion will only focus on the scenario where the confinement scale, $\Lambda_H$, is smaller than $m_\chi$, such that non-perturbative physics is less important. Our model is also viable for larger $\Lambda_H$, and we briefly discuss this possibility. In Sec.~\ref{sec:pheno}, we present the predictions for direct detection and proton decay signals, finding that current constraints require DM masses of at least $10^7$ GeV. We also calculate the hidden-sector contributions to the Higgs potential, finding an improved stability of the electroweak vacuum. We turn to the cosmology of such heavy DM in Sec.~\ref{sec:cosmology}, where we discuss DM production by inflaton decay in Sec.~\ref{sec:DM_production_inflaton_decay} and production from the SM thermal bath in Sec.~\ref{sec:thermal_relic}.  We summarize our findings and comment on future directions in Sec.~\ref{sec:conclusion}. An additional mechanism for DM production during reheating is discussed in Appendix~\ref{app:DM_production_inelastic}.

\section{Gauge Coupling Unification}
\label{sec:coupling}

\begin{figure}[tbhp]
	\centering
	\includegraphics[width=0.45\linewidth]{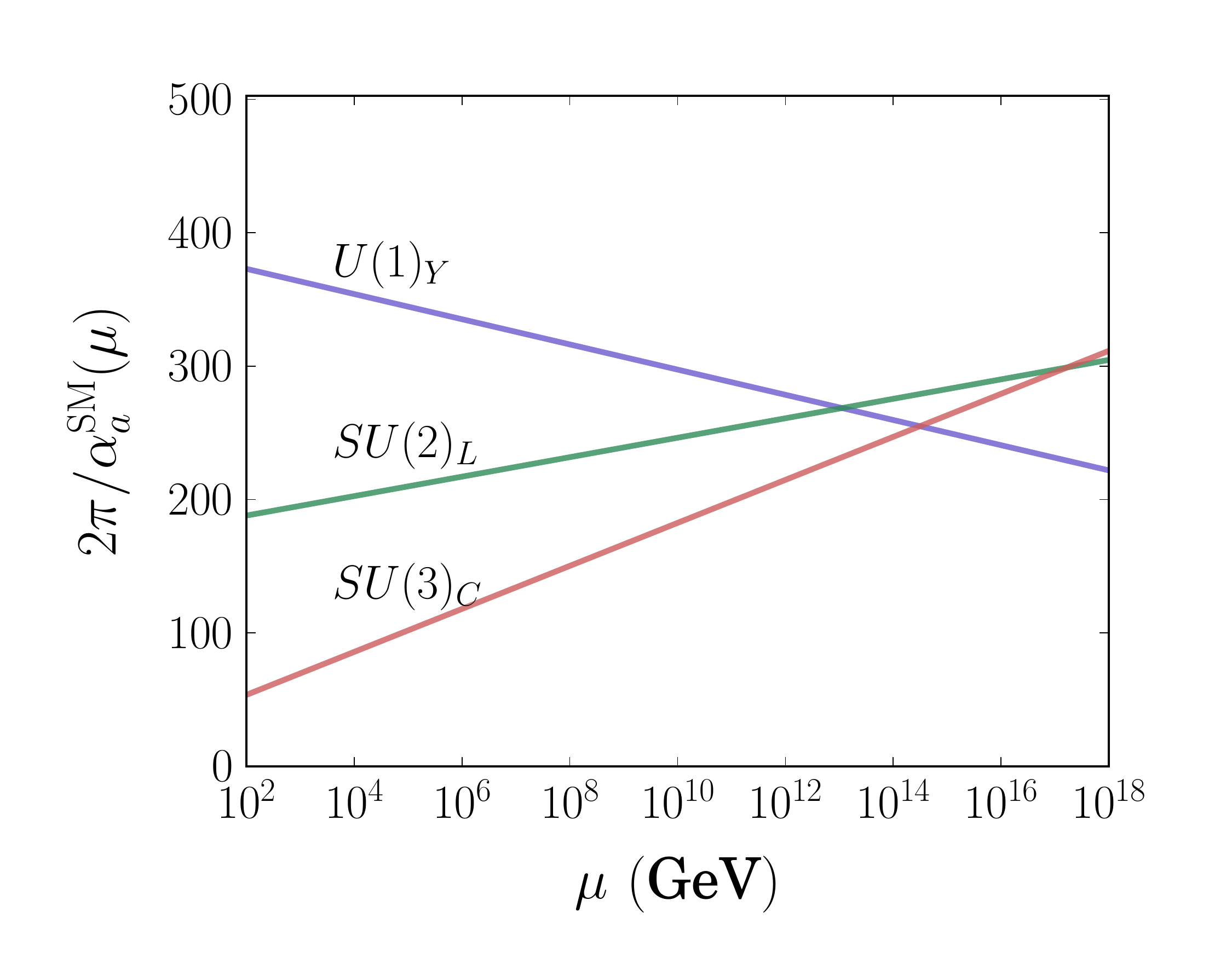}
	\includegraphics[width=0.45\linewidth]{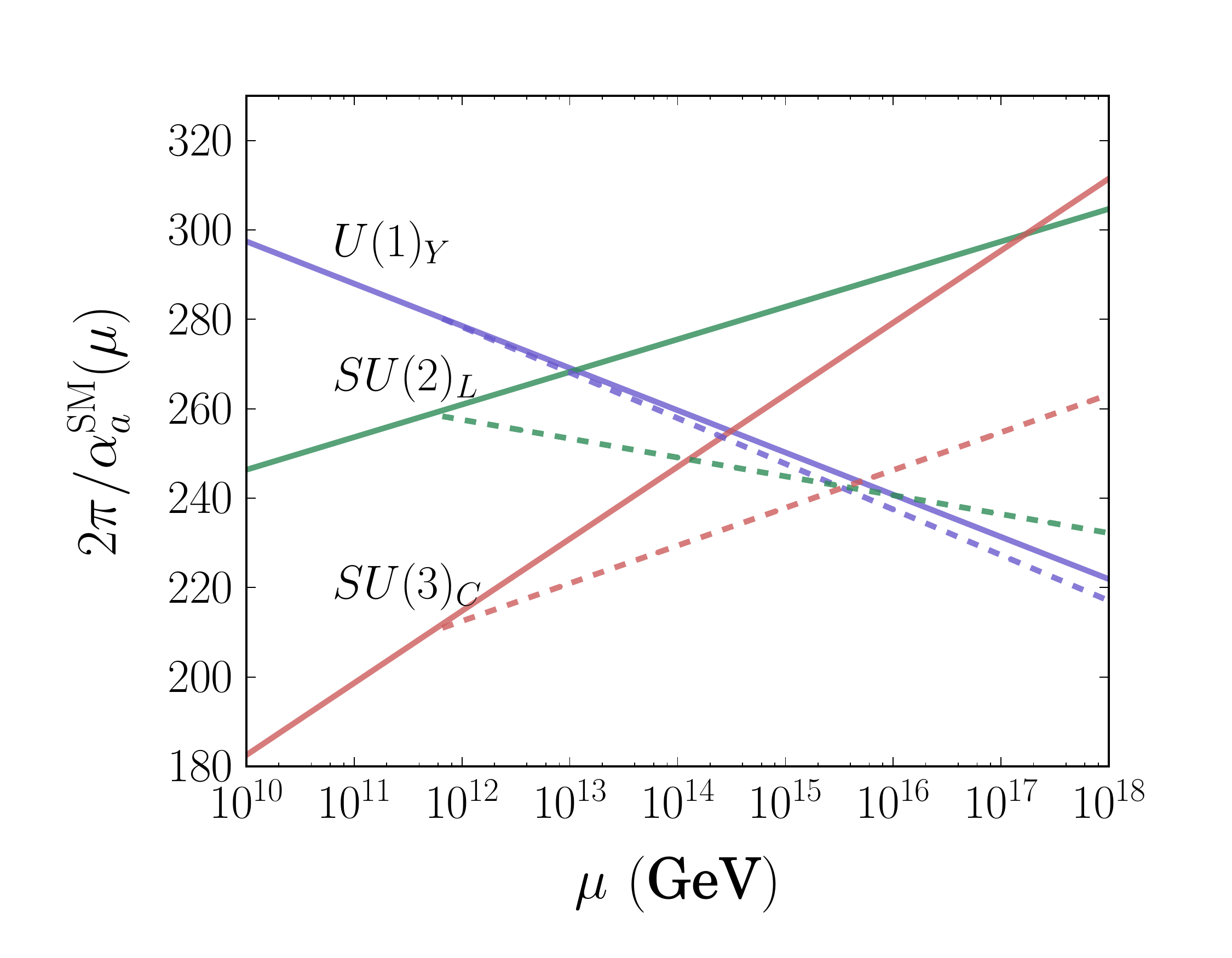}
	\caption{\sl \small (Left) Running of the gauge coupling constants in the standard model. (Right) Running of our benchmark GUTzilla model with 3 pairs of Dirac fermions transforming as $({\bf 3, 2})_{\frac{1}{6}}$.
}
\label{fig:run_SM}
\end{figure}

The SM has gauge group $SU(3)_C \times SU(2)_L \times U(1)_Y$, where the coupling strength for each of these gauge interactions receives scale-dependent quantum corrections. At one loop, they are given by%
\footnote{The hypercharge coupling strength is defined as $\alpha_1 = \sfrac{5 g_Y^2}{12\pi}$, where the extra factor of $\sfrac{5}{3}$ comes from embedding $U(1)_Y \subset SU(5)$.}
\begin{align}
	\label{eq:renormalization}
	\frac{2\pi}{\alpha^\SM_a(\mu)} = \frac{2\pi}{\alpha_a(m_Z)} + b_a^{\SM} \log \frac{\mu}{m_Z},
\end{align}
where $b_a^{\SM}$ is the running coefficient in the SM, 
\begin{align}
	b_1^{\SM} = - \frac{41}{10}, \quad
	b_2^{\SM} = \frac{19}{6}, \quad
	b_3^{\SM}=7.
\end{align}
Fig.~\ref{fig:run_SM} illustrates the SM running coupling at one-loop. The three couplings approach one another at around $\mu \sim 10^{15}$ GeV but do not unify. Gauge coupling unification then requires new matter lighter than $\sim 10^{15}$ GeV or a huge threshold correction around the unification scale. 

The simplest way to achieve unification is by adding a new matter field $\chi$ with mass $m_{\chi} \lesssim 10^{15}$ GeV, as illustrated in the right panel of Fig.~\ref{fig:run_SM}. The addition of $\chi$ modifies the running couplings in Eq.~\ref{eq:renormalization} as
\begin{align}
	\label{eq:renormalization_mod}
	\frac{2\pi}{\alpha_a(\mu)} = 
	\begin{cases}
		\frac{2\pi}{\alpha^\SM_a(\mu)} 
		&\quad   \mu < m_\chi,  \\
		\frac{2\pi}{\alpha^\SM_a(\mu)} 
		 + b_a^{\chi} \log \frac{\mu}{m_\chi}
		&\quad \mu \geq m_\chi.
	\end{cases}
\end{align}
The coefficient $b_a^\chi$ can be written as $b_a^\chi = -N_\chi s_\chi c_a$, where $N_\chi$ is the overall multiplicity of the $\chi$ multiplet, $c_a$ is the sum of the Dynkin index, and $s_\chi$ is the spin factor defined as
\begin{align}
  s_\chi = \frac{1}{3} \times
\left\{
  \begin{array}{clccl}
    \sfrac{1}{2} & (\text{real scalar}), && 2 & (\text{Weyl fermion}),\\
    1 & (\text{complex scalar}), && 4 & (\text{Dirac fermion}).
  \end{array}\right.\,
\end{align}
Unification requires that all three couplings meet%
\footnote{
We do not consider the strong-coupling unification scenario~\cite{Maiani:1977cg,Goldberger:2002pc,Dermisek:2012as} nor the unification involving higher-dimensional operators~\cite{Hill:1983xh,Shafi:1983gz}.
}
at some scale $\Lambda_\GUT$, i.e. ${{\alpha_a(\Lambda_\GUT)=\alpha_\GUT}}$. Given that the gauge coupling unification is only dependent on the combination $b^\chi_a \log\frac{m_\chi}{\Lambda_\GUT}$ at one loop, $\Lambda_\GUT$ and $\alpha(\Lambda_\GUT)$ are therefore invariant under the transformation, 
\begin{align}
  	N_\chi s_\chi \rightarrow n\, N_\chi s_\chi \qquad
  	m_\chi \rightarrow m_\chi \left(\frac{\Lambda_\GUT}{m_\chi} \right)^{1-\frac{1}{n}}\,,
	\label{eq:GUT_multiplicity_transform}
\end{align}
where $n$ is an arbitrary constant. Using the transformation in Eq.~\ref{eq:GUT_multiplicity_transform}, $m_\chi$ can be raised arbitrarily close to $\Lambda_\GUT$. 

More generally, since $\chi$ is part of a unified multiplet, the  conditions for unification will be modified depending on the mass scale of the multiplet.
Let $\chi \oplus \chi'$ be a complete GUT multiplet and assume that the masses of $\chi'$ are all of the same order, $m_{\chi'}$.
The requirement of coupling unification using Eq.~\ref{eq:renormalization_mod} implicitly assumes that $m_{\chi'}=\Lambda_\GUT$, which can be relaxed. For fermionic $\chi$, $m_{\chi'}$ is naturally below $\Lambda_\GUT$. Above the scale $m_{\chi'}$, the extra running from $\chi\oplus \chi'$ comes in a complete multiplet and does not affect unification. Gauge coupling unification then depends only on $b^\chi_a \log\frac{m_\chi}{m_{\chi'}}$ at one loop, and thus will be invariant under the following set of transformations:
\begin{align}
  	N_\chi s_\chi &\rightarrow n\, N_\chi s_\chi \qquad
 	 m_\chi \rightarrow m_\chi \left(\frac{m_{\chi'}}{m_\chi} \right)^{1-\frac{1}{n}}
	\qquad {\rm and } 
	\label{eq:GUT_general_multiplicity_transform}
	\\
	m_\chi &\rightarrow \kappa m_\chi
	\qquad \;\;\;
	m_{\chi'} \rightarrow \kappa m_{\chi'}.
	\label{eq:GUT_scale_transform}
\end{align}
The transformations in Eq.~\ref{eq:GUT_general_multiplicity_transform} preserve both $\alpha(\Lambda_\GUT)$ and $\Lambda_\GUT$. However,  Eq.~\ref{eq:GUT_scale_transform} keeps $\Lambda_\GUT$ the same, but modifies $\alpha(\Lambda_\GUT)$ due to the extra running between $(m_{\chi'},\Lambda_\GUT)$. So while the full running is a function of $(m_\chi, m_{\chi'}, N_\chi s_\chi)$, the transformations above demonstrate that $\Lambda_\GUT$ is solely dependent on the SM representation of $\chi$ at one loop.

A preliminary analysis of allowed $\Lambda_\GUT$ can thus immediately place restrictions on allowed representations for $\chi$. Perturbatively, the presence of additional matter generally causes the gauge coupling to run larger. If the $\chi$ multiplet is $SU(2)_L$ neutral, then $2\pi/\alpha_1$ and $2\pi/\alpha_3$ are pushed to smaller values; comparing with Fig.~\ref{fig:run_SM}, we see that no solution with unification is possible in this case. Next, consider a $\chi$ multiplet which is $SU(3)_C$ neutral: while unification is now possible, the maximum possible $\Lambda_\GUT$ is where $\alpha_1^{\SM}$ and $\alpha_3^{\SM}$ intersect, or $\Lambda_\GUT < 3.4\times 10^{14}$~GeV. However, $\Lambda_\GUT$ directly controls the proton decay rate, and current bounds require a GUT scale of at least $\mathcal{O}(10^{15})$ GeV~\cite{Nishino:2012bnw,Takhistov:2016eqm}. We conclude that $\chi$ must be charged with respect to both $SU(3)_C$ and $SU(2)_L$.\footnote{\setstretch{0.8}We can waive this conclusion if the masses of $\chi'$ are also split. See e.g.~Refs.~\cite{Ibe:2009gt,Aizawa:2014iea}. We do not consider this possibility as it requires more tuning. }

How does a color-charged $\chi$ give rise to a DM candidate?
Without additional structure, the color-charged $\chi$ particles will form bound states with light quarks, resulting in strong interactions with ordinary matter. Such strongly interacting DM has already been ruled out by earth heating and direct detection bounds~\cite{Mack:2007xj,Mack:2012ju,Akerib:2015rjg}. On the other hand, the DM-nucleon interaction will be suppressed if the Bohr radius of the bound state is much smaller than $1/\Lambda_{\QCD}$, which can be achieved by adding a new confining gauge group $G_H$ with a large confinement scale $\Lambda_H$. The DM is then a neutral composite bound state of $\chi$, though it generally has nonzero hypercharge. Stringent direct detection bounds due to $Z$ exchange then leads us to consider a very heavy, non-thermal DM candidate. We turn to specifics of this scenario in the following section.

\section{Minimal GUTzilla}
\label{sec:model}

In the minimal GUTzilla DM model, we add to the SM an extra Dirac fermion multiplet $\chi \oplus \chi'$, where $m_\chi < m_{\chi'}$. In Sec.~\ref{sec:coupling}, we have shown that $\chi$ needs to be charged under both $SU(3)_C$ and $SU(2)_L$. The smallest such representation for $\chi$ is $({\bf 3},{\bf 2})_{\sfrac{1}{6}}$, which is a subset of the  $\bf 10$ and $\bf 15$ representations of an $SU(5)$ GUT.%
\footnote{Achieving gauge coupling unification by adding $({\bf 3},{\bf 2})_{\sfrac{1}{6}}$ was first considered in Ref.~\cite{Murayama:1991ah}.}
(While our considerations do not depend on the unification group, we will use the language of $SU(5)$ for simplicity.)

In order to form color-neutral DM, we introduce a hidden sector gauge group $SU(N_H)$ (with $N_H = 3$), which confines at $\Lambda_H$.\footnote{\setstretch{0.8}There are other possibilities for the hidden sector gauge group, such as $SO(2N)_H$. We briefly comment on this in Sec.~\ref{sec:conclusion}.} The $\chi\oplus \chi'$ multiplet transforms as the fundamental representation of $SU(3)_H$.  Then the GUTzilla DM is a stable baryonic state composed of three $\chi$ fermions. Depending on $\Lambda_H$, the composite sector also contains new meson or glueball states, which decay quickly into SM particles.

Depending on the hierarchy between $\Lambda_H$, $m_\chi$ and $m_{\chi'}$, our previous analysis of gauge coupling unification may be modified. Our model has three distinctive physical regimes:
\begin{itemize}
	\item  $\Lambda_H  < m_\chi < m_{\chi'}$: The gauge running computation is simplest for this hierarchy, with heavy $\chi$. The SM gauge couplings receive new contributions at the scale $m_\chi$, where the hidden sector coupling is perturbative, and it is straightforward to determine the running at one-loop. The hidden baryons are composites of the $\chi$ fermions, such that $m_{\rm DM} \approx 3 m_\chi$. 
	\item $m_\chi < \Lambda_H < m_{\chi'}$: In this case, hidden sector pions $\pi_H$ are present and the SM gauge couplings are modified at the scale $m_{\pi_H}$. The running between $m_{\pi_H}$ and $\Lambda_H$ can be calculated in chiral perturbation theory. Non-perturbative physics comes in around the confinement scale and will introduce extra threshold corrections. For scales larger than $\Lambda_H$, the perturbative one-loop analysis applies again. One can estimate the correction to the running in the chiral regime. Since the SM gauge group explicitly breaks the chiral flavor symmetry, the pion masses are only smaller than $\Lambda_H$ by a loop factor. Then the change to $2\pi/\alpha_\SM$ is at most of order $|\log(\alpha_\SM/4\pi)| \lesssim 5$. Given the small running coefficient due to scalars, such a contribution is subdominant to potential threshold corrections near $\Lambda_\GUT$. We will then treat this scenario in the same way as the heavy-$\chi$ case, keeping in mind that the running calculation applies with the substitution $m_{\chi} \rightarrow \Lambda_H$ and a large uncertainty exists from extra running due to pions and other non-perturbative composite states.
	\item $m_\chi < m_{\chi'} < \Lambda_H $: The changes to the SM gauge coupling running mainly arise from the light pions in the hidden sector. Again, the pion masses have large contributions due to SM gauge interactions and are one-loop suppressed compared to $\Lambda_H$. The resulting pion spectrum has small mass splittings, and thus the modification of the SM running is too small to achieve unification. 
\end{itemize}

For concreteness, we discuss below the case where $\Lambda_H  < m_\chi < m_{\chi'}$, such that we may follow the perturbative one-loop analysis.  As noted above, for $m_\chi < \Lambda_H < m_{\chi'}$, the gauge coupling unification is very similar as long as we make the identification $m_\chi \simeq \Lambda_H$ in the running calculation. For a Dirac fermion $\chi$ in the representation $({\bf 3},{\bf 2})_{\sfrac{1}{6}}$, the contribution to the running in Eq.~\ref{eq:renormalization_mod} is
\begin{align}
	\left(b_1^{\chi},b_2^{\chi},b_3^{\chi} \right) = 
		N_\chi  s_\chi \times \left( -\frac{1}{10}, -\frac{3}{2}, -1 \right).
\end{align}
Assuming coupling unification, the GUT scale is given by $\Lambda_\GUT = 3 \times 10^{15}$ GeV, and the mass hierarchy between $m_\chi, m_{\chi'}$ is given by
\begin{align}
	\log_{10} \left(\frac{m_\chi}{m_{\chi'}}\right) = - \frac{12.6}{N_\chi s_\chi }\,,
\label{eq:GUT_mchi}
\end{align}
where $N_\chi s_\chi = 3 \times 4/3$ for our model. Such a small $m_\chi/m_{\chi'}$ can be achieved by
tuning a Yukawa coupling of the $\chi \oplus \chi'$ multiplet with a GUT-breaking Higgs field.
Unification can also be achieved for scalar $\chi \oplus \chi'$, which we do not discuss here.

The confinement scale of the hidden sector $\Lambda_H$ is in general a free parameter. For example, for $\chi \oplus \chi'$ transforming as a ${\bf 10}$ multiplet of $SU(5)$, the renormalization group equation of the gauge coupling of $SU(3)_H$ is given by
\begin{align}
	\frac{\rm d}{{\rm dln}\mu} \frac{2\pi}{\alpha_H(\mu)} = \left\{
	\begin{array}{ll}
		\frac{13}{3} & ,\quad \mu > m_{\chi'} \\
		7 & , \quad m_{\chi}<\mu < m_{\chi'} \\
		11 & ,\quad \mu <m_\chi 
	\end{array}
	\right. .
\end{align}
The gauge group $SU(3)_H$ will remain asymptotically free and $\Lambda_H$ can range all the way from $\Lambda_H \ll 1$ GeV to $10^{14}$ GeV for moderate coupling at GUT scale, $\alpha_H(\Lambda_\GUT) \in (0.01, 1)$.

\subsection{GUTzilla dark matter}

For $\chi \sim ({\bf 3},{\bf 2})_{\sfrac{1}{6}}$, the lightest baryons of the hidden sector are in a SM doublet $({\bf 1, 2})_{\sfrac{1}{2}}$. Writing $\chi=(\chi_u, \chi_d)$ as a doublet of $SU(2)_L$, these composite states have wave-functions $(\DM, \DM^+)=(\chi_u\chi_d\chi_d, \,\chi_u\chi_u\chi_d)$, where the color and hidden $SU(3)_H$ indices are contracted with the antisymmetric $\epsilon$-tensor.

For the regime we are interested in, the inverse radius of the composite particle is much larger than the electroweak scale.
Then the doublet $(\DM, \DM^+)$ is essentially an elementary particle at low energies, and the DM-nucleon scattering rate is dominated by $Z$-exchange. Electroweak symmetry breaking effects will induce a mass splitting for the doublet, which is independent of $m_\DM$~\cite{Cirelli:2005uq}:
\begin{align}
m_{\DM^+} - m_{\DM} = \alpha_2 m_W \sin^2\left(\frac{\theta_W}{2}\right) \left(Q^2+\frac{2YQ}{\cos\theta_W}\right)
\simeq 340 \,\MEV.
\label{eq:DM_splitting}
\end{align}
The charged $\DM^+$ particle can decay through an off-shell $W^+$, which can lead to a soft pion or leptons. The two-body decay $\DM^+ \rightarrow \DM + \pi^+$ dominates, with a rate given by
\begin{align}
\Gamma_{\DM^+} \simeq
\frac{\pi \alpha_2^2 V_{ud}^2 f_\pi^2}{2 m^4_W}(m_{\DM^+} - m_{\DM})^3
\sqrt{1-\frac{m_\pi^2}{(m_{\DM^+} - m_{\DM})^2}} = \frac{\hbar}{1.5 \, {\rm n sec}}\,,
\end{align}
such that the $\DM^+$ easily decays away before Big-Bang Nucleosynthesis (BBN).

Meanwhile, the stability of the neutral DM state can be guaranteed by symmetries. One could simply impose a $\mathbb{Z}_2$ charge $(-1)^\chi$, or a continuous $U(1)_\chi \supset (-1)^\chi$. These symmetries can also be obtained within $SO(10)$ unification. For example, one can embed the $\chi$ within a $\bf{45}$ or ${\bf 54}$ multiplet of $SO(10)$, so that $\chi$ has a $U(1)_{B-L}$ charge of $2/3$. When $SO(10)$ is broken into $SU(5) \times U(1)_{X}$ by a ${\bf 126}$ Higgs, a discrete $(-1)^{3(B-L)}$ remains unbroken. Then $(-1)^\chi$ can be identified with $(-1)^{3(B-L)+F}$, where $F$ is fermion number. Geometrically, such a parity can be thought of as the spinor parity in $SO(10)$. Analogous to the Lorentz group, representations of the Lie algebra $\mathfrak{so}(10)$ are actually representations of the universal cover ${\rm Spin}(10)$, where ${\rm Spin}(10)/\mathbb{Z}_2 = SO(10)$, and all spinor representations (SM fermions) are charged under the extra $\mathbb{Z}_2$. Then $(-1)^\chi$ can be identified with $\mathbb{Z}_2 \times (-1)^F$.

In addition to the GUTzilla DM, there are additional composite states arising in the hidden sector. The physics of these states depend on $\Lambda_H$ and $m_\chi$. While our main focus is on the heavy-$\chi$ scenario ($\Lambda_H < m_\chi$), we will also discuss the alternative QCD-like case ($\Lambda_H > m_\chi$) for completeness. Both scenarios provide a stable DM candidate and similar low-energy phenomenology.

\subsection{Heavy-$\chi$ scenario}

When the hidden confinement scale $\Lambda_H$ is much smaller than $m_\chi$, the lightest hidden sector states are glueballs with various spin and quantum numbers \cite{Morningstar:1999rf}. There are additional heavier meson states which decay rapidly into glueballs and SM gauge particles. The lightest glueball is a scalar and can decay back into the SM through dimension-8 operators obtained from integrating out the $\chi\oplus \chi'$~\cite{Juknevich:2009ji}. These operators can be written schematically as
\begin{align}
	\mathcal{L}\supset &\frac{\alpha_\SM \alpha_H}{m_\chi^4} \bigg[
	c_1 F_\SM^2 G_H^2 + c_2 \left(F_\SM \widetilde F_\SM \right) \left(G_H \widetilde G_H\right) + \textrm{ higher spin terms}\bigg],
\end{align}
where the $c_i$ are $\mathcal{O}(1)$ coefficients.  The higher spin terms include non-trivial tensor contractions between the SM field strengths and the higher spin glueball fields. The dimension-8 operators induce decay of the scalar glueballs into SM gauge bosons, with a rate of order
\begin{align}
	\Gamma_{\rm glueball} \simeq \frac{\alpha_\SM^2 \Lambda_H }{2 \pi} 
	\left(\frac{\Lambda_H}{m_\chi}\right)^8.
\end{align}
As long as $\Lambda_H$ is sufficiently large, the glueball will decay well before BBN ($\sim 1$ sec):
\begin{align}
  \Lambda_H \gtrsim 
  50\, \TeV \,
  \bigg(  \frac{1\ {\rm sec}} {\tau} \bigg)^{\frac{1}{9}}
  \bigg( \frac{10^{-2}}{\alpha_\SM^2 } \bigg)^{\frac{1}{9}}
  \bigg( \frac{m_\chi}{10^8 \GeV} \bigg)^{\frac{8}{9}}\,.
\label{eq:quirky_bound}
\end{align}
If CP is conserved, there are additional higher spin states that can only decay radiatively, which will lead to a stronger bound for $\Lambda_H$.  Then Eq.~\ref{eq:quirky_bound} will serve as a conservative bound for the hidden sector confinement scale.

\subsection{QCD-like scenario}
When the confining scale of the hidden sector is larger than $m_\chi$, the hidden sector undergoes chiral symmetry breaking. The light degrees of freedom are pseudo-Nambu-Goldstone bosons, or pions $\pi_H$. From the perspective of the hidden sector, there is an approximate $SU(3\times 2)_L \times SU(3\times 2)_R$ global symmetry explicitly broken by SM gauge interactions. 
Below the confinement scale, this flavor symmetry is spontaneously broken to the diagonal group $SU(3\times 2)_V \supset SU(3)_C \times SU(2)_L$. There are a total of 35 pion fields, and they reside in SM representations given by ${ ({\bf 8,3})_0 \oplus({\bf 8,1})_0 \oplus ({\bf 1,3})_0}$ and with masses
\begin{align}
	m^2_{\pi_H} \sim \Lambda_H m_\chi + \frac{\alpha_\SM}{2\pi} \Lambda_H^2 .
\end{align}
The DM again is a baryon doublet, but its mass is dominated by the confinement scale $\Lambda_H$ instead of the masses of its constituents,
\begin{align}
	m_\DM \simeq N_H \Lambda_H,
\end{align}
where $N_H= 3$.

The pions can decay through dimension-5 operators in chiral perturbation theory, 
\begin{align}
\mathcal{L}_{\chi \rm PT} &\supset
\frac{N_H }{ 4\pi F_{\pi_H} } \bigg[ \sqrt{\frac{3}{5}}
\frac{\sqrt{\alpha_3 \alpha_1}}{6} \pi_{\footnotesize H}^{a}G_{\mu\nu}^a \widetilde B^{\mu \nu }+
\sqrt{\frac{3}{5}} \frac{ \sqrt{\alpha_2 \alpha_1} }{4} \pi_H^{A}W_{\mu\nu}^A \widetilde B^{\mu \nu }
 \notag \\
&\quad\quad\quad\quad \ \
 + \frac{\sqrt{\alpha_3 \alpha_2}}{4} \pi_H^{aA} G_{\mu\nu}^a \widetilde W^{\mu \nu A}
+ 
\frac{\alpha_3}{2} d^{abc} \pi_H^{a} G_{\mu\nu}^b \widetilde G^{c\mu \nu }  \bigg]
\,,
\label{eq:pion_decay}
\end{align}
which easily satisfies BBN constraints for the DM masses considered.

\section{Phenomenology}
\label{sec:pheno}

In this section, we consider the main phenomenological implications of GUTzilla DM, namely DM direct detection and proton decay. Strong constraints on direct detection experiments require a large DM mass, $m_\DM \gtrsim 10^8\ \GeV$. Additionally, we comment on the modification to the Higgs potential and vacuum stability, finding that the inclusion of the hidden sector improves stability. 

\subsection{Direct Detection}
\label{sec:direct_detection}

Given that the DM has a non-vanishing $U(1)_Y$ charge, it interacts with a nucleus via tree-level exchange of a $Z$ boson. For a given nucleus $N$, the average per-nucleon scattering cross section is given by
\begin{eqnarray}
	\sigma_{n} = \frac{G_F^2 \mu_n^2}{2\pi} Y^2 \bigg[ \frac{(A_N-Z_N) -\left(1-4 \sin \theta_W^2\right)Z_N }{A_N} \bigg]^2  \, ,
\label{eq:sigmaDMN}
\end{eqnarray}
where $G_F$ is the Fermi constant, $\mu_n$ is the reduced mass of the nucleon and DM, $Y$ is the hypercharge of DM, and $A_N$ and $Z_N$ are the atomic number and charge of the nucleus, respectively. Such an interaction is highly constrained by direct detection experiments; in the high-mass limit, the tightest bounds come from the LUX experiment~\cite{Akerib:2015rjg}:
\begin{align}
	\sigma_{n} < 10^{-44} {\rm cm}^2 \left(\frac{m_{\DM}}{1~{\TEV}}\right) 
\qquad
{\rm and }
\qquad 
	m_{\rm DM} > 
		5 \times 10^7~{\rm GeV}\times(2Y)^2 \, .
\label{eq:DMbound}
\end{align}
This constraint on the DM mass translates to constraints on the hidden sector. In the case $\Lambda_H \lesssim m_\chi$, the DM mass bound leads to a bound on $m_\chi \simeq m_\DM/3$. For $\Lambda_H \gtrsim m_\chi$, the DM mass bound leads to a bound on the hidden sector confinement scale $\Lambda_H \simeq m_\DM/N_H $. Together with gauge coupling unification, $m_\DM$ can roughly be in the range $10^{8}$ to $10^{12}$ GeV.

In Fig.~\ref{fig:proton_DD} we show the sensitivity of direct detection and proton decay experiments to our model. We show the constraint from LUX (2015)~\cite{Akerib:2015rjg} by the vertical solid line, assuming the $m_{\rm DM}= 3\times m_{\chi}$. We also show the projected sensitivity of the LZ experiment~\cite{Akerib:2015cja}, and a direct detection experiment whose sensitivity is limited by the neutrino background~\cite{Billard:2013qya}. It can be seen that a large portion of the parameter space can be tested by future direct detection experiments. With multiple target nuclei, it is also possible to test whether a DM candidate interacts via $Z$ exchange~\cite{Feldstein:2013uha}, which would point towards very heavy DM as in Eq.~\ref{eq:DMbound}.

\subsection{Proton Decay}
In a generic GUT, there are new heavy particles at the unification scale that can mediate proton decays (for a review, see Ref.~\cite{Langacker:1980js}). In an $SU(5)$ GUT, a proton can decay into a meson and a lepton via the exchange of gauge bosons charged under both $SU(3)_C\times SU(2)_L \subset SU(5)$. The most stringent proton decay constraint comes from the $p\rightarrow  \pi^0 + e^+$ channel. In the effective field theory below the GUT scale, such a decay arises from the following dimension-6 operator,
\begin{align}
	{\cal L}_{\rm int} \approx \frac{g^2_{\rm GUT}}{M_{XY}^2}\left[
	A_R \left(u_L e_L\right) \left( \bar{d}_R^\dag \bar{u}_R^\dag  \right) 
	+ 2 A_L \left( d_L u_L\right) \left( \bar{u}_R^\dag \bar{e}_R^\dag \right)
\right],
\label{eq:proton_decay_operator}
\end{align}
where $M_{XY}$ is the mass of GUT gauge bosons, $g_{\GUT}$ is the gauge coupling constant at the GUT scale, and $A_{R,L}$ is the Wilson coefficient from renormalization group running from the GUT scale down to the hadron scale. Here we have neglected the effects of the quark mixing angles since this depends on how quark masses are unified. We have also ignored potential contributions from the Higgs sector, which are Yukawa-coupling suppressed. Then the decay rate of the proton is given by
\begin{align}
	\Gamma^{-1}(p\rightarrow \pi^0 e^+ ) & =
	\left[\frac{1}{32\pi}m_p\left(1- \frac{m_{\pi^0}^2}{m_p^2}\right)^2 \frac{g_{\rm GUT}^4}{M_{XY}^4}(A_R^2+4A_L^2) |W_0|^2 \right]^{-1}  \\
	& \simeq 3.4 \times 10^{33}
	~{\rm years}~\times  
	\left( \frac{\alpha^{-1}(M_{XY})}{40} \right)^2
	\left(  \frac{M_{XY}}{3\times10^{15}~{\rm GeV}} \right)^4
	\left(  \frac{0.103~{\rm GeV}^2}{W_0} \right)^2 
\,, \nonumber
\end{align}
where $W_0$ is the quantity encoding the form factor of a pion and a proton. Lattice calculations show that $|W_0|=0.103$ GeV$^2$ at the renormalization scale of 2 GeV, with uncertainty of 40\%~\cite{Aoki:2013yxa}.

\begin{figure}[t!]
\centering
\includegraphics[width=0.65\linewidth]{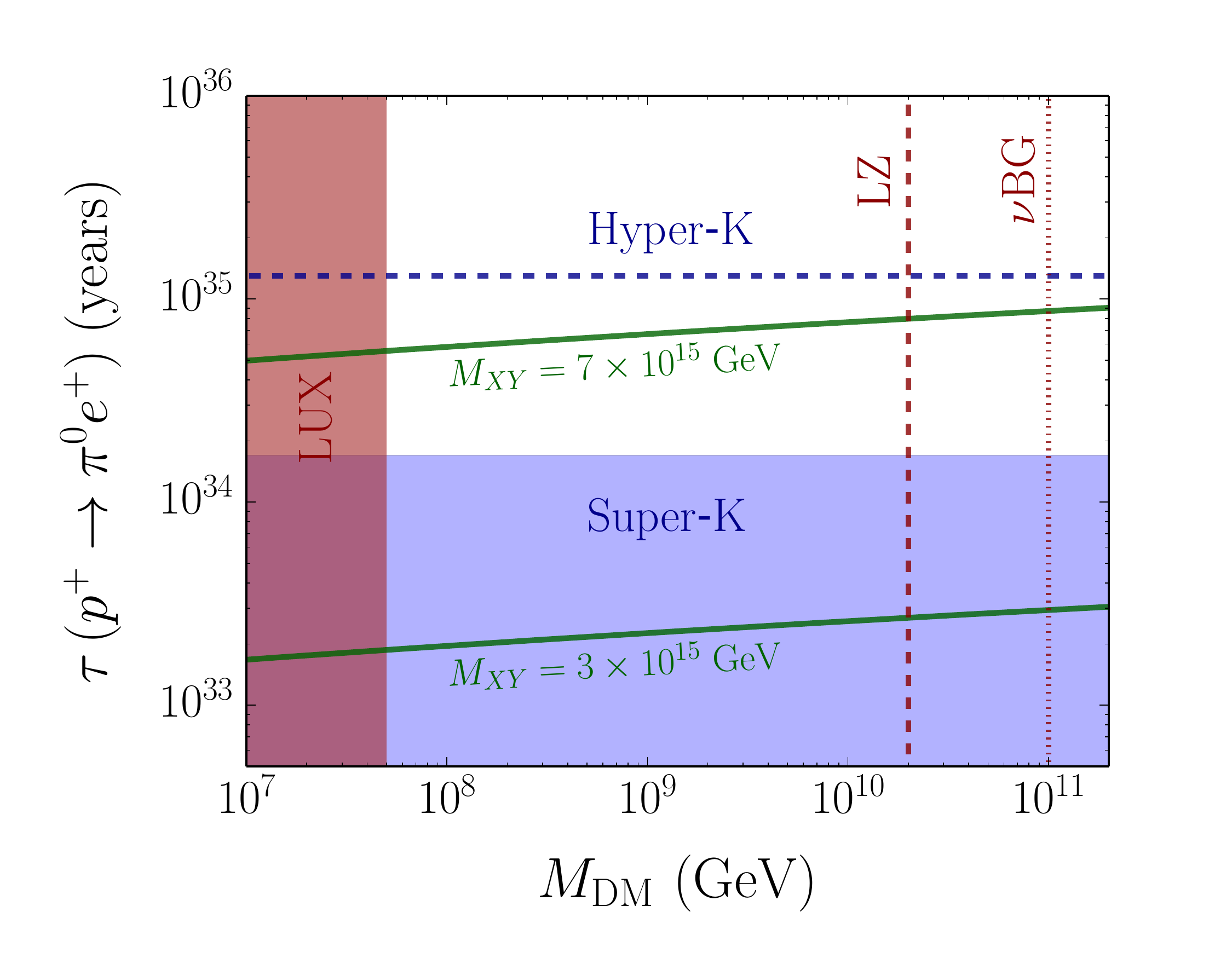}
\caption{\sl \small
Direct detection and proton decay constraints are shown for our minimal GUTzilla model.
The green lines give the prediction for the decay rate of $p \to \pi^0 e^+$ for different values of $M_{XY}$ ($3$ and $7 \times 10^{15}\; \GeV$) corresponding to the nominal and the maximal $M_{XY}$ given threshold corrections $\Delta_{\max}= 5$. The light blue shaded region is excluded by Super-Kamiokande~\cite{Takhistov:2016eqm}, while the horizontal dashed line shows the sensitivity of Hyper-Kamiokande. For proton decay lifetimes, we use the central value for $W_0 = 0.103~{\rm GeV}^2$.  The red shaded region shows the constraint from LUX~\cite{Akerib:2015rjg}, while the two dashed lines show the projected sensitivity of the LZ experiment and a direct detection experiment whose sensitivity is limited by the neutrino background.  
}
\label{fig:proton_DD}
\end{figure}

In Fig.~\ref{fig:proton_DD}, we show the prediction for the decay rate of the proton in our model as a function of $m_{\chi}$.
The lower green line shows the prediction for $M_{XY}=\Lambda_{\rm GUT}=3\times 10^{15}$ GeV.
The upper green line shows the prediction for $M_{XY}=7\times 10^{15}$ GeV, which can be achieved by a moderate threshold correction around the GUT scale, $\Delta (2\pi/\alpha)=5$, as we discuss below.
In fixing $\alpha_{\rm GUT}$, we assume that $\chi$ is embedded into a $\bf{10}$ of $SU(5)$. If $\chi$ is embedded into a larger representation, $\alpha_{\rm GUT}$ is larger and the proton decay rate becomes larger.
The light blue shaded region is excluded by Super-Kamiokande, $\Gamma^{-1}(p\rightarrow \pi^0 e^+ )>1.7\times 10^{34}$~years (90\%CL)~\cite{Takhistov:2016eqm}. We show the expected sensitivity of Hyper-Kamiokande, $\Gamma^{-1}(p\rightarrow \pi^0 e^+ )>1.3\times 10^{35}$~years (90\%CL)~\cite{Abe:2011ts}. It can be seen that the entire parameter space can be covered by Hyper-Kamiokande. The calculations for $M_{XY}, A_{L,R}$ and the treatment of threshold corrections are described below.

\paragraph{Estimation of $M_{XY}$}
The masses of the $X/Y$ gauge bosons are typically of order $\Lambda_\GUT$.
If threshold contributions to the running are present, then $M_{XY}$ can be raised and the proton lifetime can be increased. 
Generally, an accurate estimate for $M_{XY}$ requires taking into account any additional split multiplets around $\Lambda_\GUT$ and/or higher order corrections to the running couplings. In order to account for these model-dependent corrections, we simply relax the coupling unification requirement by varying the mass ratio $m_\chi/m_{\chi'}$ and allowing the couplings to differ by some amount. Thus, we can determine $M_{XY}$ by demanding that
\begin{align}
 \left| \frac{2\pi}{\alpha^a(M_{XY})} - \frac{2\pi}{\alpha^b(M_{XY})} \right| \le \Delta_{\max} \,,
\label{eq:threshold}
\end{align}
for all the SM couplings, and where $\Delta_{\max}$ parameterizes the deviation from unification.
With only the SM particle content, the minimal value of $\Delta_{\max}$ is $25$ around $M_{XY}\sim 10^{14}$ GeV,
which is ruled out.
With the additional $\chi$ multiplet, the case $\Delta_{\max} = 0$ corresponds to the scenario of no threshold corrections with $M_{XY} = \Lambda_\GUT = 3\times 10^{15}$ GeV. Allowing $\Delta_{\max} = 5$, $M_{XY}$ can be in the range $1.5 \times 10^{15}$ to $7.0\times 10^{15}$ GeV. The green curves in Fig.~\ref{fig:proton_DD} show the proton decay lifetime given two values of $M_{XY}$, corresponding to $\Delta_{\max}= 0$ and the upper bound on  $M_{XY}$ for $\Delta_{\max}=5$.

\paragraph{Estimation of $A_{R,L}$}

The dimension-6 operators in Eq.~\ref{eq:proton_decay_operator} obtain anomalous dimensions from gauge interactions. Under renormalization group evolution, the Wilson coefficients receive significant multiplicative corrections. The coefficient $A_{R,L}$ at different scales is then related by~\cite{Buras:1977yy,Wilczek:1979hc}:
\begin{align}
	A_R(\mu) =
	\left(\frac{\alpha_3(\mu)}{\alpha_3(M)}\right)^{\frac{2}{b_3}}
	\left(\frac{\alpha_2(\mu)}{\alpha_2(M)}\right)^{\frac{9}{4b_2}} 
	\left(\frac{\alpha_1(\mu)}{\alpha_1(M)}\right)^{\frac{11}{12b_1}} \times A_R(M), \nonumber \\
	A_L(\mu) =
	\left(\frac{\alpha_3(\mu)}{\alpha_3(M)}\right)^{\frac{2}{b_3}}
	\left(\frac{\alpha_2(\mu)}{\alpha_2(M)}\right)^{\frac{9}{4b_2}} 
	\left(\frac{\alpha_1(\mu)}{\alpha_1(M)}\right)^{\frac{23}{12b_1}} \times A_L(M)\,.
\end{align}
Taking $A_{L,R}(\Lambda_\GUT)=1$, and $M \simeq \Lambda_\GUT \simeq M_{XY}=10^{15}$ GeV, one obtains $A_R^{\SM}(2\, \GeV)\simeq 3.0$ and $A_L^{\SM}(2\, \GeV) \simeq 3.4$ in the SM. In our GUTzilla model, the introduction of $\chi\oplus \chi'$ only modifies $A_{L,R}$ at the percent level. Since possible threshold corrections dominate the uncertainty in the proton lifetime, we simply take $A_{L,R}$ to be the SM values in our calculation.

\begin{figure}[t!]
\centering
\includegraphics[width=0.65\linewidth]{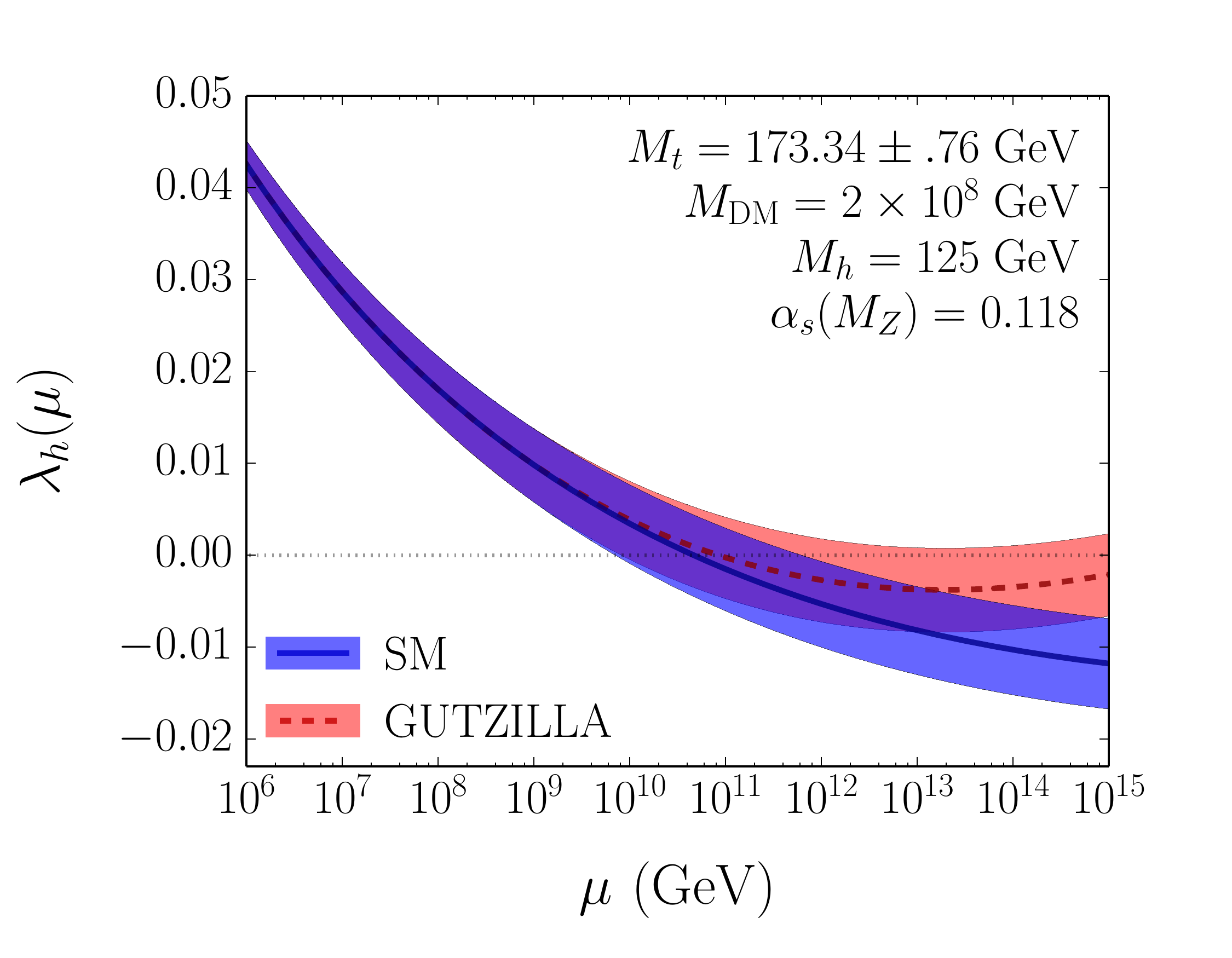}
\caption{\sl \small
Running of the SM Higgs quartic coupling with only the SM (blue, solid) and in the presence of the additional multiplets $\chi\oplus\chi'$ (red, dashed). $m_{\chi'}= 10^{11}\, \GeV$ is chosen in order to satisfy coupling unification. The largest uncertainty in the running comes from determination of the top mass; the bands shown are $\pm1\sigma$ in $M_{t}$.}
\label{fig:quartic}
\end{figure}

\subsection{Vacuum Stability}
\label{sec:vacuum}

In the SM, the Higgs quartic coupling receives a large negative contribution from the top Yukawa coupling, which can lead to a meta-stable or unstable electroweak vacuum~\cite{Casas:1994qy, Casas:1996aq,Hambye:1996wb}. Given the current Higgs mass and top mass measurements, an NNLO calculation for the Higgs potential has firmly excluded SM vacuum stability at the 2$\sigma$ level (see Ref~\cite{Degrassi:2012ry, Alekhin:2012py, Buttazzo:2013uya} and references therein). The SM Higgs quartic becomes negative at around $10^{11}$ GeV; the presence of the additional $\chi$ multiplet with $m_\chi \lesssim 10^{11}$ GeV could increase the gauge coupling and improve the stability of the Higgs potential. For our minimal GUTzilla model, a small $m_\chi \lesssim 6 \times 10^{7}$ GeV is needed to stabilize the Higgs potential within 1$\sigma$. The Higgs quartic running is illustrated in Fig.~\ref{fig:quartic}, which shows the quartic coupling including the leading-order effect of $\chi, \chi'$. We use the central value for the Higgs mass, and show the effect of varying the top-quark pole mass within $\pm1\sigma$.

\section{GUTzilla Cosmology}
\label{sec:cosmology}

A massive DM candidate in thermal equilibrium during a radiation-dominated era is easily overproduced; unitarity limits on  the DM annihilation cross section require ${m_\DM \lesssim 300\; \TEV}$ \cite{Griest:1989wd}. Instead, processes before the end of reheating can set the abundance of GUTzilla DM and thus get around this bound, as long as the reheating temperature $ T_{\RH}$ is less than the DM mass.  In this section we describe the various possibilities.

Superheavy DM may be produced gravitationally during the transition from an inflationary phase to a matter-dominated era~\cite{Chung:1998zb,Chung:2001cb,Chung:2011ck}.   This mechanism is sufficient for producing the correct relic abundance of DM if both $T_\RH$ and the Hubble scale at the end of inflation, $H_I$, are large enough. In addition, constraints on isocurvature perturbations are satisfied for DM masses $m_\DM \gtrsim 6 H_I$~\cite{Chung:2004nh}. In large-field inflation, where $H_I$ is typically as large as the inflaton mass $m_\phi$, the condition translates into $m_\DM \gtrsim m_\phi$.

\begin{figure}[t!]
\vspace{0.5cm}
\centering
\begin{minipage}{\textwidth}
    \begin{fmffile}{direct_decay}
      \setlength{\unitlength}{1cm}\large
      \begin{fmfgraph*}(9,3)
        \fmfforce{(0.2w,0.5h)}{b}
        \fmfforce{(0w,0.5h)}{l}

        \fmfforce{(0.52w,0.7h)}{s1}
        \fmfforce{(0.68w,0.8h)}{s2}
        \fmfforce{(0.84w,0.9h)}{s3}

        \fmfforce{(.63877w,-.000286h)}{c1}
        \fmfforce{(.687w,.0297h)}{c2}

        \fmfforce{(.7988w,.09971h)}{c3}
        \fmfforce{(.8468w,.12971h)}{c4}

        \fmfforce{(.9588w,.1997h)}{c5}
        \fmfforce{(1.0068w,.2297h)}{c6}

        \fmfforce{(1w,1h)}{sm}
        \fmfforce{(0.3w,-0.2h)}{x1}
        \fmfforce{(0.32w,-0.18h)}{x2}
        \fmfforce{(0.34w,-0.16h)}{x3}
        
        \fmf{dashes,label=$\phi$}{l,b}

        \fmfblob{.05w}{b}
        \fmf{plain}{b,x1}        
        \fmf{plain}{b,x2}
        \fmf{plain}{b,x3}
        
        \fmf{dbl_plain_arrow, label.dist=20, label.side=left, label=${\rm SM}$}{b,sm}

        \fmf{curly, tension=3}{s1,v1}
        \small
        \fmf{plain, label.dist=10, label.side=right, label=${}_{\bar\chi}$}{c1,v1} 
        \fmf{plain, label.dist=10, label.side=right, label=${}_\chi$}{v1,c2} 

        \fmf{curly, tension=3}{s2,v2}
        \fmf{plain, label.dist=10, label.side=right, label=${}_{\bar\chi}$}{c3,v2} 
        \fmf{plain, label.dist=10, label.side=right, label=${}_\chi$}{v2,c4} 

        \fmf{curly, tension=3}{s3,v3}
        \fmf{plain, label.dist=10, label.side=right, label=${}_{\bar\chi}$}{c5,v3} 
        \fmf{plain, label.dist=10, label.side=right, label=${}_\chi$}{v3,c6} 
       
      \end{fmfgraph*}
    \end{fmffile}
\end{minipage}

\vspace{1.5cm}
\caption{\sl  \label{fig:inflaton_decay}
DM production by  decay of the inflaton $\phi$, as described in Sec.~\ref{sec:DM_production_inflaton_decay}. The process is dominated by soft gauge boson emissions from SM charged particles and their subsequent splitting into $\bar\chi\chi$ pairs.}
\end{figure}
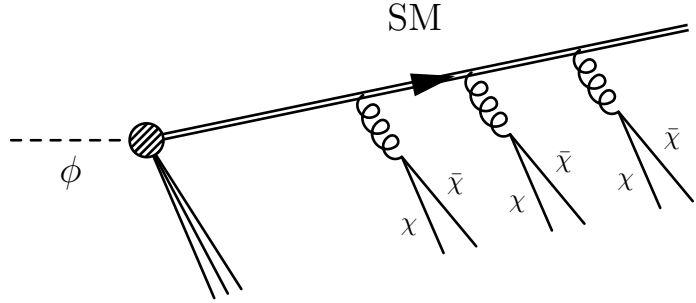

If $m_\DM < m_\phi$, production during reheating is also possible~\cite{Kolb:1998ki, Chung:1998rq, Chung:2001cb, Feldstein:2013uha, Allahverdi:2002nb, Allahverdi:2002pu, Kurata:2012nf, Harigaya:2014waa}. There are three possible mechanisms in play: inflaton decay, thermal production, and inelastic scattering between inflaton decay products and the SM plasma. DM production from inflaton decay, shown in Fig.~\ref{fig:inflaton_decay}, will be important as long as it is kinematically accessible. For heavy GUTzilla DM with $m_\DM \gtrsim 10^8$ GeV, overproduction of DM will then place constraints on the reheating temperature~\cite{Harigaya:2014waa}, which we will discuss in Sec.~\ref{sec:DM_production_inflaton_decay}. However, these constraints can be evaded if the inflaton dominantly decays to SM singlets which are lighter than the DM, thus shutting off the previous mechanism. Then other possibilities such as thermal production and/or inelastic scattering of inflaton decay products will become important. We will discuss the thermal production channel in Sec.~\ref{sec:thermal_relic}. The inelastic scattering case is highly model-dependent, and we give a simplified treatment in Appendix~\ref{app:DM_production_inelastic}.

In each of these scenarios, GUTzilla DM production is further complicated by the hidden sector dynamics. If the temperature of the hidden sector at the time of DM production is smaller than $\Lambda_H$, GUTzilla DM will be directly produced. Otherwise, the constituents $\chi$ will first be produced and the DM bound states are formed only after the hidden sector confining phase transition.
As before, we will primarily focus on the heavy-$\chi$ scenario, $m_\DM \gg \Lambda_H$. Note that in the QCD-like scenario, non-perturbative processes 
produce a substantial amount of DM, leading to a strong constraint on $T_\RH$ unless $m_\DM > m_\phi$.

\begin{figure}[t!]
\centering
\includegraphics[width=0.65\linewidth]{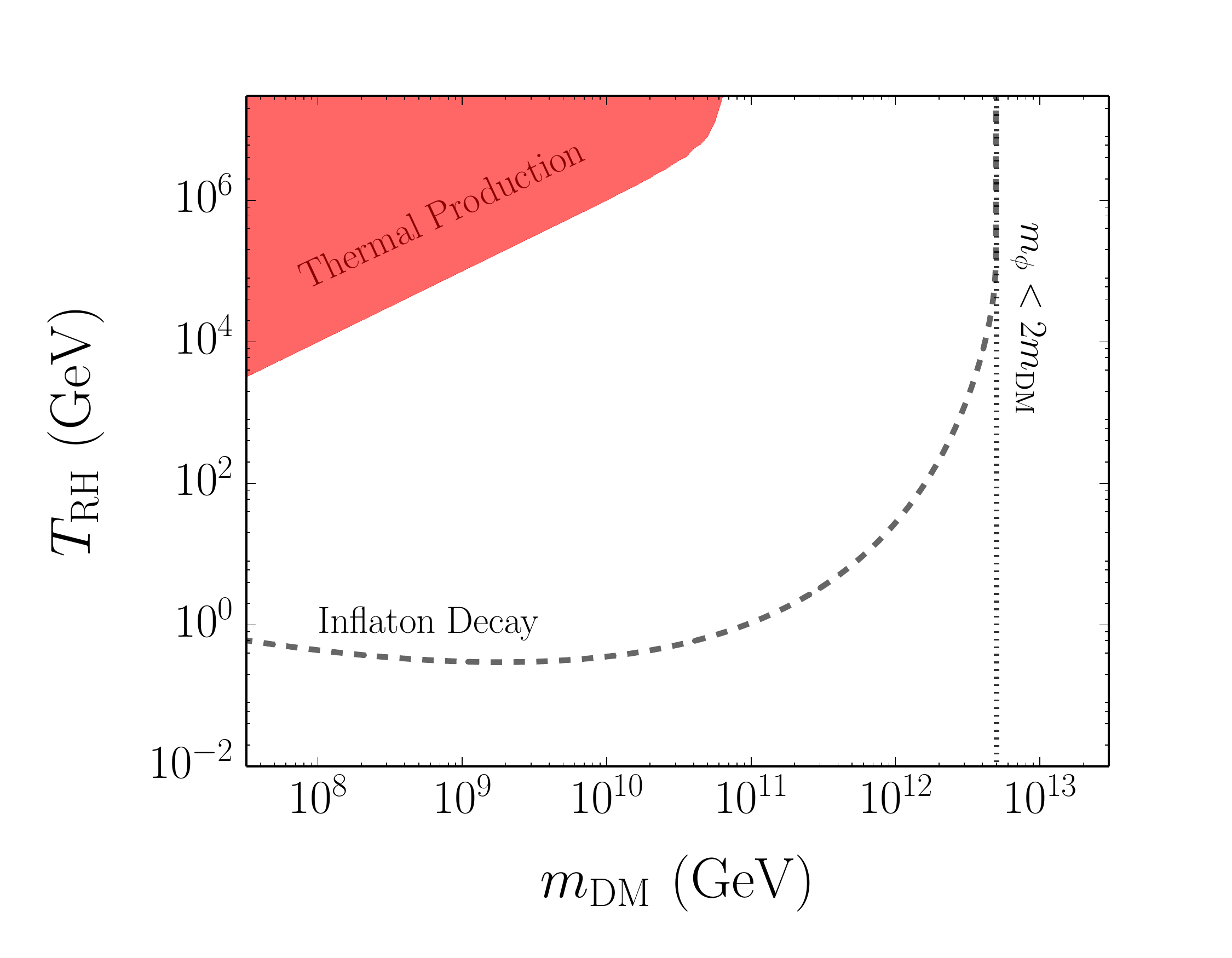}
\caption{\sl \small
Bounds on the reheating temperature from DM production, with inflaton mass $m_\phi = 10^{13}$ GeV. The dashed-black line shows the upper limit on $T_{RH}$ coming from direct decays of the inflaton (Sec.~\ref{sec:DM_production_inflaton_decay}). When direct decays of the inflaton to DM are turned off, the thermal production of DM during reheating can give the dominant relic abundance (Sec.~\ref{sec:thermal_relic}) -- in this case, the red-shaded region indicates where $\Omega_{\DM} h^2$ is greater than the observed value. For the constraint from direct decays, Eq.~\ref{eq:DM_frag} is used for an estimate of the average DM multiplicity $\langle N_\DM \rangle$ in inflaton decay, $\alpha$ is fixed at 0.05, and we include an additional factor of $(1-4m_\DM^2/m_\phi^2)^{1/2}$ to take into account phase-space suppressions.
}
\label{fig:reheat_temp}
\end{figure}

Fig.~\ref{fig:reheat_temp} illustrates the parameter space to produce GUTzilla DM, depending on the production mechanism, $T_\RH$, and $m_\DM$. The derivation of these bounds can be found in the remainder of this section.
For direct inflaton decay, a low $T_\RH$ is required in order to avoid over-producing DM; this constraint is shown as the dotted-black line.  If the inflaton only couples to SM singlets, and the singlet has suppressed coupling to SM and hidden sector states, a looser constraint from thermal production applies and is shown in the red region. In this case, the correct DM relic abundance can readily be obtained for high $T_\RH$. In addition, the constraints can be evaded when DM becomes heavy enough such that it is kinematically inaccessible. In the direct decay case, this happens when $2m_\DM > m_\phi$, shown as the dashed black line.

\subsection{Inflaton Decay}
\label{sec:DM_production_inflaton_decay}

When the inflaton directly decays to SM-charged particles, production of DM can proceed by gauge boson emissions and subsequent splitting into $\chi$ particles.\footnote{Showering of heavy SM-charged particles from the decay of an inflaton is discussed in Ref~\cite{Kurata:2012nf} in the context of the minimal supersymmetric SM.} An illustrative diagram is shown in Fig.~\ref{fig:inflaton_decay}.
Due to the large entropy production during the reheating period, DM production from inflaton decay is most prominent at the end of reheating, around $T=T_\RH$. The DM relic density from inflaton decay can be estimated as
\begin{align}
	\frac{\rho_\DM}{s} \simeq \frac{m_\DM T_\RH}{m_\phi}\langle N_\DM\rangle
\end{align}
where $T_\RH$ is the reheating temperature, $m_\phi$ is the inflaton mass, and  $\langle N_\DM\rangle$ is the average number of composite baryonic DM per inflaton decay. Generally, $\langle N_\DM\rangle$ depends on the inflaton coupling. We will take the most conservative approach and assume there is no direct coupling of the inflaton with $\chi$. However, as long as the inflaton primarily decays into SM charged particles, the decay products can undergo showering and radiate hidden sector particles, which can eventually hadronize into DM. At high energies, these showering processes are perturbative and can be calculated systematically~\cite{Furmanski:1979jx}. 

Consider the average number of $\chi$ particles produced in the shower. A $\bar\chi\chi$ splitting is necessarily preceded by a gauge boson emission. Then at leading order, the average number of $\chi$  from an inflaton decay is
\begin{align}
  \langle N_\chi\rangle = \sum_a \frac{\alpha^a}{6\pi}\int_{4m_\chi^2}^{m_\phi^2}\frac{dk^2}{k^2} N^a_{\rm gauge}(k^2),
\label{eq:chi_frag}
\end{align}
where $N^a_{\rm gauge}$ is a splitting kernel given by
\begin{align}
  N^a_{\rm gauge}(k^2) \simeq \frac{\alpha^a}{2\pi}C^a_F \log^2 \left(\frac{m_\phi^2}{k^2} \right)\,,
\end{align}
and  $a$ denotes SM and hidden sector gauge bosons. $N_{\rm gauge}^a$ is the average number of gauge bosons at leading order, and $\alpha^a$ and $C_F^a$ are the corresponding gauge coupling and Casimir for gauge boson emission. Note the form of $N^a_{\rm gauge}$ is valid only for a non-Abelian gauge group, which is assumed to dominate the shower. Performing the integration in Eq.~\ref{eq:chi_frag} gives
\begin{align}
  \langle N_\chi \rangle \sim \sum_a \frac{C^a_F \alpha^{a2}}{36\pi^2}\log^3\left(\frac{m_\phi^2}{4m_\chi^2}\right)\,.
\end{align}
For $m_\phi \gg m_\chi$, resummation of the large logarithm results in an exponential enhancement~\cite{Furmanski:1979jx}.
We find that as long as $m_\phi^2/4m_\chi^2 \lesssim 10^{10}$, the perturbative estimate in Eq.~\ref{eq:chi_frag} is sufficient. 

For GUTzilla models, the DM is composed of three $\chi$, which will require three separate gauge boson splittings to $\bar\chi\chi$  in addition to a 
suppression factor to form a 
baryon, which we take to be $\sim 1/N_H^2$.  The average DM multiplicity is then estimated to be
\begin{align}
  \langle N_\DM \rangle \sim \frac{1}{N_H^2}\left[ \frac{C^a_F \alpha^{a2}}{36\pi^2}\log^3\left(\frac{m_\phi^2}{4m_\chi^2}\right)\right]^3 \,.
\label{eq:DM_frag}
\end{align}
For a typical SM interaction, $\alpha\sim 0.05$, and for $m_\phi/m_\chi \sim 10^{5}$, we have ${\langle N_\DM \rangle \sim 10^{-4}}$. 

So far we have only included perturbative contributions from showering and Eq.~\ref{eq:DM_frag} ignores contributions from non-perturbative processes, which is valid in the heavy-$\chi$ limit. In the opposite QCD-like regime where $m_\chi/\Lambda_H$ is small, non-perturbation fragmentation and hadronization can also produce baryons, leading to a large DM multiplicity. In the Lund string model, baryon fragmentation can be thought of as breaking of the gluon string by a diquark/anti-diquark pair, with a fragmentation function of the form $\sim \exp(-4m_\chi^2/{\Lambda_H^2})$~\cite{Andersson:1983ia, Andersson:1984af}. In the light quark regime, diquark fragmentation is not significantly suppressed and the DM multiplicity will be of order the hidden gluon multiplicity $\langle N_\DM \rangle\sim 2\alpha \log^2(m_\phi^2/4m_\chi^2)/27\pi \sim 0.01$. In the heavy-$\chi$ limit $m_\chi \gg \Lambda_H$, diquark fragmentation is exponentially suppressed and does not contribute to baryon production.

Given our very conservative estimate of $\langle N_\DM\rangle$, in order to avoid overproducing the DM relic density ${\rho_\DM/s \lesssim 10^{-9}}$ GeV, we find the reheating temperature is constrained to be
\begin{align}
  T_\RH \lesssim 1\,\GeV \,
\bigg(\frac{10^8 \, \GeV}{m_\DM }\bigg)
\bigg(\frac{m_\phi}{10^{13}\, \GeV}\bigg)
\bigg(\frac{10^{-4}}{\langle N_\DM\rangle}\bigg)\,.
\end{align}
The black dashed curve in Fig.~\ref{fig:reheat_temp} shows the regions of parameter space excluded from overproduction of the DM, where Eq.~\ref{eq:DM_frag} is used for an estimate of the average DM multiplicity from inflaton decay. We see that even a suppressed $\langle N_\DM\rangle$ can lead to tight constraints on the reheating temperature. The constraint can be relaxed, however, if no direct coupling between the inflaton and SM charged particles exists; then sub-dominant processes become important, as we discuss below and in Appendix~\ref{app:DM_production_inelastic}.

\subsection{Thermal Production}
\label{sec:thermal_relic}

If the inflaton only couples to SM singlets $S$ (e.g., a right-handed neutrino), the decay into DM will have to proceed through the coupling between $S$ and other SM particles, which can be highly suppressed. Subsequent decay from $S \rightarrow $ DM can be forbidden as long as $m_S < m_\DM $. Then production of DM from the SM thermal bath is relevant, and the right relic density can be achieved thanks to dilution from entropy production by the inflaton. 

During reheating, the inflaton gradually transfers energy to the SM plasma. The SM bath will heat up to a maximum temperature $T_{\max} > T_{\RH}$, while the energy density of the universe is still dominated by that of the inflaton.
The energy density then becomes dominated by the relativistic SM bath at $T_{\RH}$. As long as $m_\chi < T_{\max}$, then the $\chi$s can be pair produced from the SM thermal bath via gauge interactions. The comoving number density of $\chi$ freezes out when $T \sim m_\chi$, and they can later be bound up into DM baryons when the temperature drops below $\Lambda_H$.

For light DM, the $\chi$ particles are in thermal equilibrium during the inflaton-dominated era. 
Following Ref.~\cite{Giudice:2000ex}, the DM relic density is given by
\begin{align}
\Omega_\DM h^2 \simeq 10^{-11} \frac{1}{N_H^2}
\frac{x_f \, \GeV^{-2}}{\langle \sigma v\rangle} \left(\frac{T_\RH}{T_f}\right)^3\,,
\end{align}
where $x_f = m_\chi /T_f$, and $T_f$ is the freeze-out temperature.
The annihilation cross section of $\chi$ is $\langle \sigma v\rangle \simeq  4\pi \alpha^2/m_\chi^2$, and we have
\begin{align}
  x_f \simeq 10 + \log\left[ \bigg(\frac{T_\RH}{80 \; \TeV}\bigg)^2 \bigg(\frac{10^8\; \GeV}{m_\DM}\bigg)^3 \bigg(\frac{x_f}{10}\bigg)^{\frac{5}{2}} \right].
\end{align}
To avoid over-production of DM, the reheating temperature is bounded by
\begin{align}
  T_\RH \lesssim 80 \;\TeV\; \bigg( \frac{m_\DM}{10^8 \; \GeV}\bigg)^{\frac{1}{3}}
  \bigg(\frac{10}{x_f}\bigg)^{\frac{4}{3}}.
  \label{eq:Tup_thermal}
\end{align}

The red shaded region in Fig.~\ref{fig:reheat_temp} shows the reheat temperatures excluded for thermal production.
When the DM mass is larger than $\sim 5 \times 10^{10}$ GeV, $\chi$ is not in chemical equilibrium. Then out-of-equilibrium production and inelastic scattering (Appendix~\ref{app:DM_production_inelastic}) processes may contribute to the DM abundance; since these are much more model-dependent, we have not shown these constraints.  Lastly, here we have assumed that $T_{\rm max}$ is always larger than $m_{\DM}$ and that kinetic equilibrium is established. A simple estimate gives $T_{\rm max} \sim (m_\phi \mpl)^{1/4} T_\RH^{1/2}$~\cite{Giudice:2000ex}, which is well above the DM mass for the parameter space shown here. However, a more detailed recent analysis shows that thermalization can be slower, with a lower $T_{\rm max} \sim \alpha_{\rm SM}^{4/5} m_\phi \left(  T_{\rm RH}^2 \mpl / m_\phi^3 \right)^{2/5}$~\cite{Harigaya:2013vwa}; depending on the specifics of this thermalization process, we expect the excluded region will be modified somewhat.

\section{Conclusion}
\label{sec:conclusion}

In this paper, we investigated a new class of models linking gauge coupling unification and DM through the introduction of a single multiplet. In order to achieve unification, the new multiplet must be a split GUT multiplet and the lighter component must be $SU(3)_C$ and $SU(2)_L$ charged. This prompted us to include a hidden confining sector to screen these interactions and leads to a composite baryonic DM. These DM can be very heavy and thus evade direct detection constraints, and provides a new motivation for considering the heavy DM WIMPzilla scenario. We refer to this as the GUTzilla DM scenario.

We presented a minimal implementation of GUTzilla DM by adding a split Dirac fermion $\chi\oplus \chi'$ multiplet, where $\chi$ transforms as $({\bf 3},{\bf 2})_{\sfrac{1}{6}}$ under the SM gauge group and as a $\bf 3$ under an $SU(3)_H$ hidden gauge group. The DM is then a baryon state made of three $\chi$. While our considerations do not explicitly depend on the hierarchy between $\Lambda_H$ and $m_\chi$, we focused on the heavy-$\chi$ case ($\Lambda_H < m_\chi$) for simplicity.

Phenomenologically, the most prominent signatures of GUTzilla DM are direct detection and proton decay. The current direct detection limit points to a GUTzilla DM with masses at least of order $10^{8}$ GeV, which will be readily tested at future LZ and Hyper-Kamiokande experiments. We also show that the addition of GUTzilla DM can improve the stability of the Higgs potential to within 1$\sigma$ as long as the DM is not too heavy.

The relic abundance of the DM is set before the end of reheating.
For DM mass larger than the Hubble scale at the end of the inflation,
the abundance is saturated by gravitational production if both the Hubble scale and the reheating temperature are large enough. 
For DM mass smaller than the inflaton mass,  production of DM during reheating is possible and can put a tight constraint on the reheating temperature.
If the inflaton directly decays into SM charged particles, DM is easily overproduced unless the reheating temperature is very low.
In the heavy-$\chi$ scenario we are considering, suppression of baryon production helps to alleviate these constraints.
On the other hand, we show that a large reheating temperature is still possible if the inflaton decays to SM singlets, assuming these singlets do not have large direct coupling to the hidden sector.
In this case, the DM abundance is saturated by thermal production during reheating.

There are many variations on the minimal GUTzilla DM that could be considered. In this paper, we introduced an $SU(3)$ confining gauge group to obtain an electromagnetic and color neutral baryon from $\chi$'s. One possibility is to introduce an $SO(2N)$ gauge group, where $\chi$ is in a fundamental representation. The lightest baryon is expected to be composed of $\chi^N$ and $\chi^{\dag N}$, and would be neutral under the SM gauge group. The DM can then be much lighter than the GUTzilla mass range.
However, a possible problem of this model is an existence of a meson composed of two $\chi$, which may be stable due to accidental $\chi$ number conservation and hence cause cosmological problems. This problem could be avoided by introducing a higher dimensional operator breaking the accidental $\chi$ number conservation. We defer further discussion of this model to future work.

Finally, one may ask whether the addition of split multiplets for gauge coupling unification introduces additional fine-tuning. This could be addressed by the anthropic principle~\cite{Barrow:1988yia,Weinberg:1987dv}, by attributing the fine-tuning $m_\chi/m_{\chi'}\ll1$ to the necessity of obtaining enough DM for structure formation~\cite{Hellerman:2005yi,Tegmark:2005dy}.
For a fixed reheating temperature and an inflaton mass, $m_\chi$ should be small enough to obtain the DM density.
The mass splitting is explained if $m_{\chi'}$ is biased toward a high energy scale. Note that any further mass splitting within $\chi'$ is disfavored, as it requires unnecessary fine-tuning as far as the DM abundance is concerned.
From the landscape point of view~\cite{Linde:1986fd,Bousso:2000xa,Susskind:2003kw},
this explanation of the splitting requires  that there is no habitable vacuum with a less fine-tuned parameter set.
To put it the other way around, if GUTzilla DM is present in our universe, we may
infer restrictions on the landscape of parameters related with the abundance of DM,
e.g.~the inflaton mass, the reheating temperature, and the decay constant of a QCD axion~\cite{Peccei:1977hh,Weinberg:1977ma,Wilczek:1977pj}.

\newpage

\section*{Acknowledgements}

We thank Daniele Bertolini, Matthew Low, Hitoshi Murayama, and Yasunori Nomura for helpful discussions,
and Timothy Cohen and Mariangela Lisanti for commenting on the draft.
This work was supported in part by the Department of Energy, Office of Science, Office of High Energy Physics, under contract No. DE-AC02-05CH11231, by the National Science Foundation under grants PHY-1316783 and PHY-1521446, and by the World Premier International Research Center Initiative (WPI), MEXT, Japan.

\appendix

\section{Dark Matter Production by Inelastic Scattering}
\label{app:DM_production_inelastic}

When the inflaton dominantly decays into SM singlets $S$,
the decay of the inflaton into DM may be suppressed.
In this case, the production of DM via inelastic scattering of $S$ decay products on the SM thermal bath could play an important role. Below, we present a simple estimate for these processes.

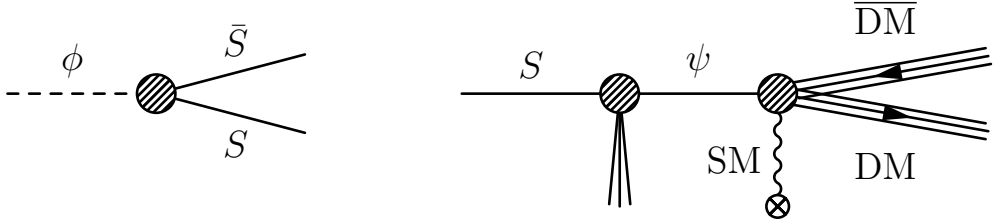
\begin{figure}[t!]
\centering
\begin{minipage}{.45\textwidth}
  \centering
    \begin{fmffile}{inflaton_singlet}
      \setlength{\unitlength}{1cm}\large
      \begin{fmfgraph*}(4,3)
        \fmfleft{i}
        \fmfright{om,o1,o2,op}
        \fmf{dashes, tension=2, label=$\phi$}{i,v}
        \fmfblob{.17h}{v}
        \fmf{plain, label.side=right, label=$S$}{v,o1}
        \fmf{plain, label.side=right, label=$\bar S$}{o2,v}
      \end{fmfgraph*}
    \end{fmffile}
\end{minipage}
\begin{minipage}{.45\textwidth}
  \centering
    \begin{fmffile}{inelastic_scattering}
      \fmfcmd{
        path quadrant, q[], otimes;
        quadrant = (0, 0) -- (0.5, 0) & quartercircle & (0, 0.5) -- (0, 0);
        for i=1 upto 4: q[i] = quadrant rotated (45 + 90*i); endfor
        otimes = q[1] & q[2] & q[3] & q[4] -- cycle;
      }
      \fmfwizard
      \setlength{\unitlength}{1cm}\large
      \begin{fmfgraph*}(7,3)
        \fmfstraight
        \fmfforce{(0w,0.5h)}{i}
        \fmfforce{(0.3w,0.5h)}{v1}
        \fmfforce{(0.6w,0.5h)}{v2}
        \fmfforce{(0.3w,0h)}{o}
        \fmfforce{(0.28w,0.01h)}{om}
        \fmfforce{(0.32w,0.01h)}{op}
        \fmfforce{(0.6w,0h)}{d1}
        \fmfright{dp,d2,d3,dp1}
        
        \fmfv{d.sh=otimes,d.f=empty, d.size=.1h}{d1}

        \fmf{plain, tension=1, label.side=left, label=$S$}{i,v1}
        \fmfblob{.17h}{v1}
        \fmf{plain, tension=1}{v1,o}
        \fmf{plain, tension=1}{v1,om}
        \fmf{plain, tension=1}{v1,op}
        \fmf{plain, tension=1, label.side=left, label=$\psi$}{v1,v2}
        \fmfblob{.17h}{v2}

        \fmf{boson, tension=1, label=${\rm SM}$}{v2,d1}

        \fmf{fermion, label.side=right, label.dist=15, label=${\rm DM}$}{v2,d2}
        \fmf{fermion, label.side=right, label.dist=15, label=$\overline{\rm DM}$}{d3,v2}

        \fmffreeze
        \fmfi{plain}{vpath (__v2,__d2) shifted (thick*(-.5,1.5))}
        \fmfi{plain}{vpath (__v2,__d2) shifted (thick*(-.5,-1.5))}

        \fmfi{plain}{vpath (__d3,__v2) shifted (thick*(-.5,1.5))}
        \fmfi{plain}{vpath (__d3,__v2) shifted (thick*(.5,-1.5))}

      \end{fmfgraph*}
    \end{fmffile}
\end{minipage}
\vspace{10mm}
\caption{\sl (left) The inflaton decays into SM singlets $S$ which are lighter than $m_\DM$. (right) $S$ decays to high energy SM-charged particles $\psi$, which scatter inelastically on the thermal hot bath to produce DM.}
\label{fig:feynman_direct_decay}
\end{figure}

The process is depicted in Fig.~\ref{fig:feynman_direct_decay}.  While the decay of $S \rightarrow $ DM is forbidden as long as $m_S < m_\DM $, the eventual decay products of $S$ must have SM charges and have energy on the order of $m_\phi$. 
Let us denote these high-energy SM charged particles as $\psi$. As reheating proceeds, $\psi$ will decay or radiate, 
and DM production can proceed through interactions between $\psi$ and the SM plasma~\cite{Harigaya:2014waa}. This is possible if the average center-of-mass energy for the interactions is large, i.e. $E_\psi T \gtrsim m_\DM^2$.
The total DM number density produced through inelastic scattering can be computed by solving the Boltzmann equation,
\begin{align}
	\frac{dn_\DM}{dt} = - 3 H n_\DM + \langle \sigma v\rangle\, n_{\SM}\, n_{\psi}
	\label{eq:DM_boltz}
	\; .
\end{align}
Here $n_{\psi}$ is the number density of the $\psi$, $n_{\SM}\sim g_* T^3$ is the number density of the SM hot bath, and $\langle\sigma v \rangle$ is the cross-section for  inelastic scattering of $\psi$ on the SM plasma. As long as the source term $\langle \sigma v\rangle\, n_{\SM}\, n_{\psi}$ is sizable, the DM density will roughly track the steady state solution in Eq.~\ref{eq:DM_boltz},
\begin{align}
	n_\DM \simeq \frac{\langle \sigma v\rangle\, n_{\SM}\, n_{\psi}}{3H} \,. 
	\label{eq:nDM_inelastic}
\end{align}

The number density of $\psi$ particles, $n_{\psi}$, depends on two competing effects: inflaton (or singlet $S$) decay and bremsstrahlung. Inflaton decay directly replenishes $n_{\psi}$, while bremsstrahlung causes hard splittings of the high energy $\psi$ particles and converts them into softer particles. 
These effects are roughly captured by the Boltzmann equation,
\begin{align}
\frac{d n_{\psi}}{dt}
\simeq  - (3H + \Gamma_{\rm split}) n_\psi
+ \frac{m_\phi}{E_\psi}\Gamma_S n_{S}
\,, 
\label{eq:nI}
\end{align}
where $n_{S}$ is the number density of $S$, and $\Gamma_{\rm split}$ is the rate of hard splitting for the $\psi$,
and $\Gamma_S$ is the decay rate of $S$ including the Lorentz boost factor.
 An extra factor $m_\phi/E_\psi$ is included as a rough estimate of the multiplicity factor. In the limit that $\Gamma_S \gg H, \Gamma_\phi$,  the number density of $S$ will reach an equilibrium density with $\Gamma_S n_S \sim \Gamma_\phi n_\phi $. Effectively, one can ignore the intermediate $S$ state and treat the inflaton as a source for production of $\psi$. A steady state solution will be reached with $n_{\psi} \simeq m_\phi \Gamma_\phi n_{\phi} / (E_\psi\Gamma_{\rm split})$ (for $\Gamma_{\rm split} \gg H$). Taking coherence effects~\cite{Landau:1953um,Migdal:1956tc} into account, the splitting rate roughly follows $\Gamma_{\rm split} \sim \alpha^2 \sqrt{T^3/E_\psi}$~\cite{Kurkela:2011ti,Harigaya:2013vwa}.

To obtain the DM density at a temperature $T$, we use Eq.~\ref{eq:nDM_inelastic} and substitute in $n_\SM \simeq g_* T^3$ for the thermal bath. For the cross section to produce DM, we take $\langle \sigma v\rangle \sim \alpha^2 \langle N_\DM( \hat s) \rangle  /(E_\psi T) $, where $\langle N_\DM( \hat s) \rangle$ denotes the average DM multiplicity per inelastic scattering event at $\hat s = E_\psi T$ (see Eq.~\ref{eq:DM_frag}).  Note that this is valid for $T < \Lambda_H$, such that the baryons are directly produced in the collision. Assuming that DM is produced primarily at a single temperature $T$, we then rescale the number density at $T$ to that at $T_\RH$, below which the comoving DM density freezes out. 

Keeping only the leading power-law dependence, the resulting DM abundance is given by
\begin{align}
	\frac{\rho_\DM}{s} \bigg|_{T_\RH} \sim \frac{ m_\DM}{N_H^2} \left[ \frac{C_F \alpha^{2}}{36\pi^2}\log^3\left(\frac{E_\psi T}{4m_\chi^2}\right)\right]^3 \frac{1}{(E_\psi T)^{3/2}} \frac{T_{\rm RH}^5}{T^2}.
\label{eq:rhoDM_inelastic}
\end{align}
In general, the full DM production must be integrated over the allowed range of $T \in [ T_{RH} , T_{max} ]$ and the allowed energy range of  $E_\psi \in  [ m_\DM^2/T , m_\phi ] $. 
Note that due to the log-enhancement in $\langle N_\DM( \hat s) \rangle$ favoring larger $E_\psi$, the production rate typically peaks at an intermediate energy. Maximizing over $T$ and $E_\psi$ in Eq.~\ref{eq:rhoDM_inelastic} to estimate the DM relic abundance, we have found that the inelastic scattering gives similar or somewhat lower relic abundance compared to thermal production. While the thermal production mechanism suffers from a larger entropy dilution, the inelastic scattering has a large suppression from $\langle N_\DM \rangle \lesssim 10^{-10}$. 

\newpage

\bibliographystyle{utphys}
\bibliography{GUT_DM}

\end{document}